# Evaluation of Mean Shift, ComBat, and CycleGAN for Harmonizing Brain Connectivity Matrices Across Sites

Hanliang Xu[a], Nancy R. Newlin[a], Michael E. Kim[a], Chenyu Gao[b], Praitayini Kanakaraj[a], Aravind R. Krishnan[b], Lucas W. Remedios[a], Nazirah Mohd Khairi[b], Kimberly Pechman[c], Derek Archer[c,d,e], Timothy J. Hohman[c,d,e], Angela L. Jefferson[c,d,f], The BIOCARD Study Team[1], Ivana Isgum[h], Yuankai Huo[a,b], Daniel Moyer[a], Kurt G. Schilling[b,i], Bennett A. Landman[a,b,d,i,j]

[a]Department of Computer Science, Vanderbilt University, Nashville, TN, USA; [b]Department of Electrical and Computer Engineering, Vanderbilt University, Nashville, TN, USA; [c]Vanderbilt Memory and Alzheimer's Center, Vanderbilt University Medical Center, Nashville, TN, USA; [d]Department of Neurology, Vanderbilt University Medical Center, Nashville, TN, USA; [e]Vanderbilt Genetics Institute, Vanderbilt University School of Medicine, Nashville, TN, USA; [f]Department of Medicine, Vanderbilt University Medical Center, Nashville, TN, USA; [g]Departments of Neurology, Johns Hopkins School of Medicine, Baltimore, MD, USA; [h]Department of Biomedical Engineering and Physics & Radiology and Nuclear Medicine, University Medical Center Amsterdam, University of Amsterdam, Amsterdam, the Netherlands; [i]Vanderbilt University Institute of Imaging Science, Vanderbilt University, Nashville, TN, USA; [j]Department of Biomedical Engineering, Vanderbilt University, Nashville, TN, USA

[1]Data used in preparation of this article were derived from BIOCARD study data, supported by grant U19–AG033655 from the National Institute on Aging. The BIOCARD study team did not participate in the analysis or writing of this report, however, they contributed to the design and implementation of the study. A listing of BIOCARD investigators may be accessed at: https://www.biocard-se.org/public/Core%20Groups.html

## ABSTRACT

Connectivity matrices derived from diffusion MRI (dMRI) provide an interpretable and generalizable way of understanding the human brain connectome. However, dMRI suffers from inter-site and between-scanner variation, which impedes analysis across datasets to improve robustness and reproducibility of results. To evaluate different harmonization approaches on connectivity matrices, we compared graph measures derived from these matrices before and after applying three harmonization techniques: mean shift, ComBat, and CycleGAN. The sample comprises 168 age-matched, sex-matched normal subjects from two studies: the Vanderbilt Memory and Aging Project (VMAP) and the Biomarkers of Cognitive Decline Among Normal Individuals (BIOCARD). First, we plotted the graph measures and used coefficient of variation (CoV) and the Mann-Whitney U test to evaluate different methods' effectiveness in removing site effects on the matrices and the derived graph measures. ComBat effectively eliminated site effects for global efficiency and modularity and outperformed the other two methods. However, all methods exhibited poor performance when harmonizing average betweenness centrality. Second, we tested whether our harmonization methods preserved correlations between age and graph measures. All methods except for CycleGAN in one direction improved correlations between age and global efficiency and between age and modularity from insignificant to significant with p-values less than 0.05.

## INTRODUCTION

Tractography derived from diffusion MRI (dMRI) offers a non-invasive, in-vivo method for identifying and measuring anatomical connections in the human brain.[1] Connectivity in tractography is commonly represented using graphs, or equivalently, two-dimensional square connectivity matrices.[2] In graph representation, nodes equate to anatomical units obtained by segmentation results of tractography, with edges representing white matter tracts connecting pairs of these units. In matrix form, rows and columns correspond to nodes, while the entry at [i, j] denotes the connection strength between node i and node j. For instance, a network consisting of n nodes and n² edges would be translated into an *n-by-n* connectivity matrix. A connectivity matrix can be characterized by its basic properties, integration, segregation, centrality, and resilience with graph theory measures.[3] By deriving certain graph measures, we gain insights into the connections between brain regions, which could then be used to understand the underlying biology of aging or diseases.[4–7]

Acquisition-related differences such as variations in scanners, scanning protocols, and reconstruction algorithms create inconsistencies among dMRI data from various sites.[8] These differences subsequently affect the multi-site analysis of the derived connectivity matrices.[9] For instance, substantial differences can be observed in the averages of connectivity matrices weighted by mean streamline length from two age-matched, sex-matched, and normal cohorts (Figure 1 and Table 1). These inconsistencies limit the potential benefits of multi-site dMRI and connectivity matrix analysis.[10] Potential benefits of harmonization include larger sample sizes, increased statistical power, and enhanced reproducibility and generalizability of results.[10]

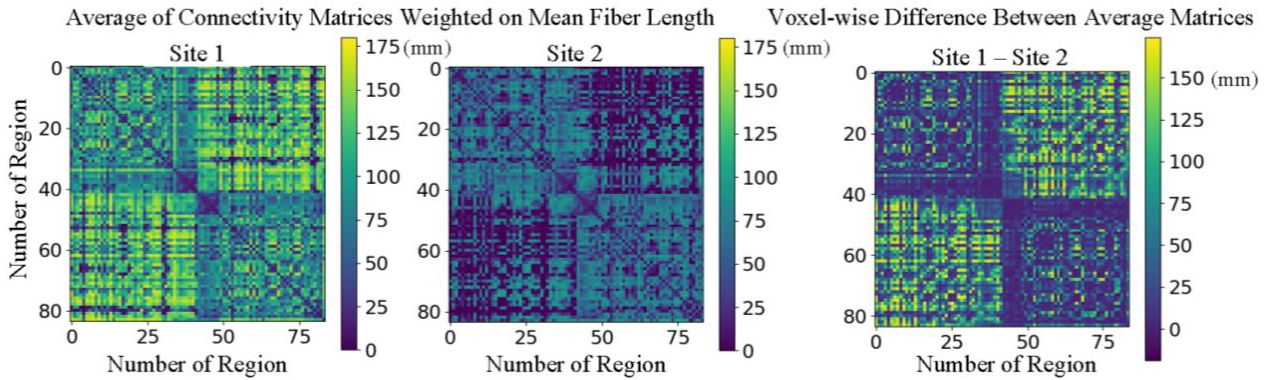

**Figure 1.** Systematic variability of connectivity matrices is high across sites. The difference matrix indicates that site 1 generates tractograms with generally longer streamlines. Note the substantial differences in the first and third quadrant. Site 2 has fewer, shorter inter-hemispheric streamlines; site 1 has more longer streamlines across hemispheres.

**Table 1.** The differences in the average connectivity matrices exist because of scanner differences introduced in the dMRI images. Harmonization at the matrix level would allow us to perform multi-site connectivity matrices analysis without the need for image-level harmonization.

|  | Modularity Mean (SD) | Global efficiency Mean (SD) |
| --- | --- | --- |
| Site 1 | 0.627 (0.0156) | 0.018 (0.0036) |
| Site 2 | 0.692 (0.0142) | 0.012 (0.0016) |

Numerous studies have attempted to harmonize dMRI.[11] Much effort has focused on harmonizing the signal before fitting.[12] For example, the rotational invariance spherical harmonics (RISH) match the energy at each rotational harmonic level across scanners.[13–15] Alternatively, some studies have focused on scanner physics and sought to calibrate the gradient nonlinearities.[16,17] Recently, there has been substantial effort to harmonize scalar measures using ComBat, a harmonization method based on an empirical Bayes framework.[18] While harmonization is typically applied on images[11], efforts to harmonize connectivity matrices have been sparse. One study applies ComBat on connectivity matrices weighted by average FA.[19] The researchers indicate that implementing ComBat on global parameters is more cost-efficient and yields superior results compared with applying it on connectivity matrices.[19] We seek to verify their findings on connectivity matrices weighted by number of streamlines and mean streamline length and evaluate two more methods of matrix harmonization.

Inspired by RISH, ComBat, and progress in the field of computer vision, we pursued three types of potential harmonization methods. First, we developed the technique of mean shift, which is analogous to RISH.[13] The core concept is simply to identify systematic average distances between sites. Second, we explored a Python implementation of ComBat[20] called neuroCombat[21, 19] and applied this directly and naively on the connectivity matrices. Finally, we explore a generative adversarial network (GAN)[22] based approach known as CycleGAN[23] that has been commonly used in the computer vision task of unsupervised image-to-image translation. This model captures the characteristics of one image domain and learns to transfer them stylistically into another.[23] It has been widely used in medical imaging to perform image harmonization and style transfer between different types of images.[24,25] Here, we evaluate these three harmonization methods. Each of these methods is newly applied in the space of connectivity matrices and could be considered a novel contribution. However, rather than emphasizing novelty, our objective is to deepen understanding of the differences between datasets at the level of connectivity matrices and to evaluate the effectiveness of current tools in harmonizing these differences.

## METHODS

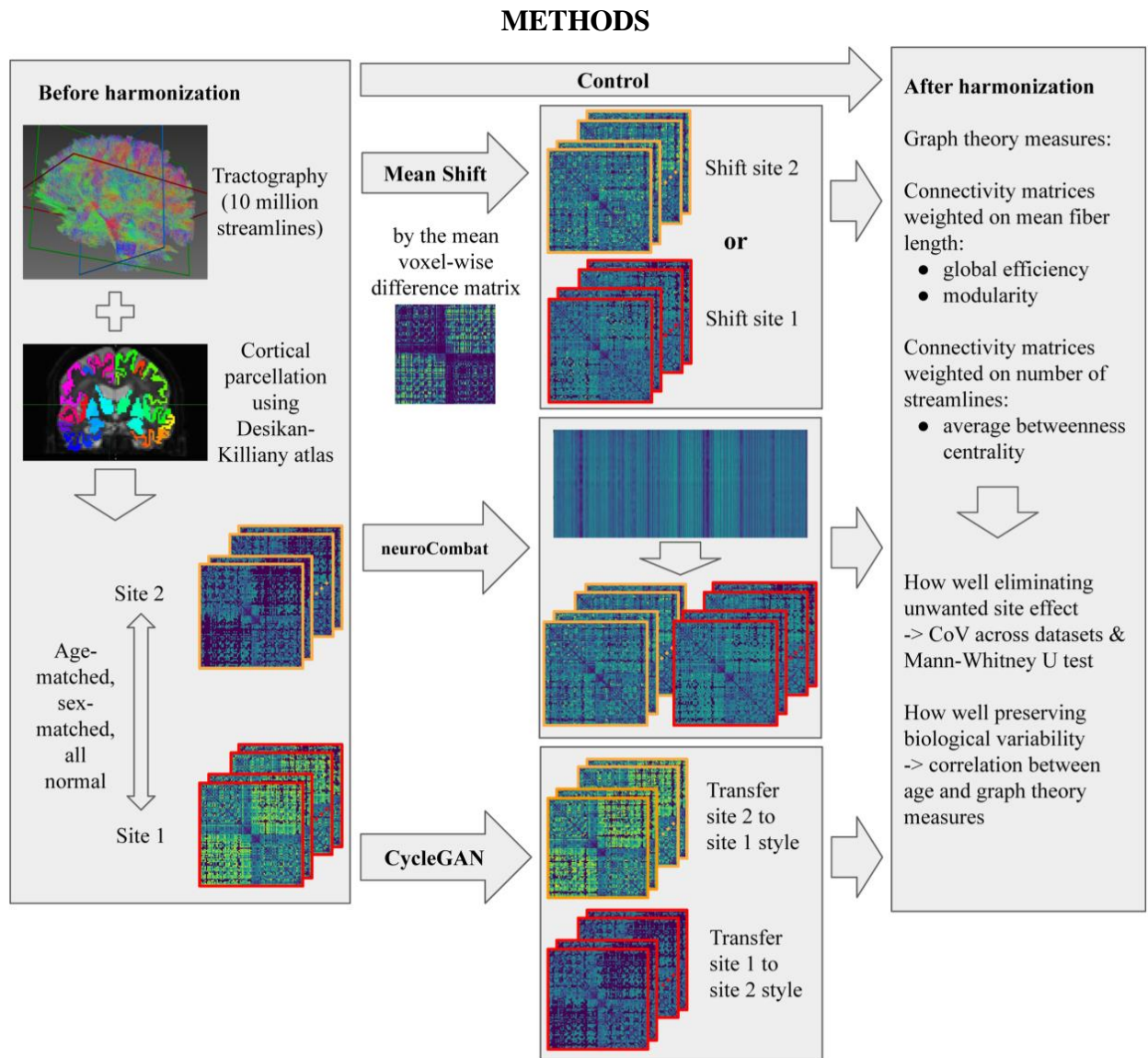

**Figure 2.** We applied identical processing to two age-matched, sex-matched, normal cohorts. We noticed significant differences of matrices and their derived graph metrics in terms of site and applied three harmonization approaches to each of the connectivity matrices directly. After harmonization, we computed graph measures based on these matrices and evaluated the methods' performance by their removal of site effect and preservation of biological variability.

Two dMRI studies were analyzed for this research: the Vanderbilt Memory and Aging Project (VMAP)[26] and the Biomarkers of Cognitive Decline Among Normal Individuals (BIOCARD)[27]. To deemphasize the specific datasets, we will refer to VMAP as "site 1" and BIOCARD as "site 2." We matched normal participants for age and sex, yielding 84 participants in each study (Table 2). The two sites' dMRI acquisition procedures differ in image resolution and b-value (Table 3). The distinction in isotropic and anisotropic resolution between the two sites is noteworthy, which could potentially explain the differences observed in cross-hemispheric streamlines.[28,29]

**Table 2.** Demographic information for participants at each site. We implemented a greedy algorithm to individually pair a normal subject in site 1 with a normal counterpart in site 2, ensuring both are of the same sex and have an age difference of no greater than one year. Given the similarity in demographic information and the size of the subject groups, it is reasonable to hypothesize that the ground truth distributions of the graph measures for these datasets would be similar to each other.

|  | Participants n | Sex n (%) female | Cognitive status n (%) normal | Age M (SD) years | Age range (median) |
|---|---|---|---|---|---|
| Site 1 | 84 | 42 (50) | 84 (100) | 72.44 (6.04) | 60-89 (72) |
| Site 2 | 84 | 42 (50) | 84 (100) | 72.49 (6.02) | 60-89 (72) |

**Table 3.** dMRI acquisition equipment and parameters at each site. The two sites are different in their b-values and scanner resolutions. The anisotropic voxel of site 2 is especially noteworthy. Studies have shown that running tractography on anisotropic voxels cause streamline loss and underestimated FA.[28,29] This could potentially explain the substantive differences in the length and amount of inter-hemispheric streamlines discussed earlier.

|  | Scanner | Resolution | Directions | b-value |
|---|---|---|---|---|
| Site 1 | Philips 3T | 2 x 2 x 2 mm$^3$ | 32 | 0 and 1000 s/mm$^2$ |
| Site 2 | Philips 3T | 0.828 x 0.828 x 2.2 mm$^3$ resampled from 2.2 x 2.2 x 2.2 mm$^3$ | 32 | 0 and 700 s/mm$^2$ |

### 2.1. Image processing

After visual quality checks of raw dMRI images, we preprocessed the images with PreQual[30]. This involved image denoising, inter-scan intensity normalization, susceptibility-, eddy current-, and motion-induced artifact correction, and slice-wise signal drop-out imputation.[30] Tractography was performed using MRtrix[31] default probabilistic tracking algorithm of Second-order Integration over Fiber Orientation Distributions with Anatomically-Constrained Tractography framework (seeding on grey matter-white matter interface, allowing backtracking, terminating using the five-tissue-type mask, generating 10 million streamlines, other parameters default). The "Desikan-Killiany" atlas[32] was utilized to define 84 nodes in the diffusion space for each participant through Freesurfer (version 6.0.0) cortical parcellations[33].

### 2.2. Connectivity matrices construction and graph measures calculation

MRtrix[31] was used to generate two 84 x 84 adjacency matrices (matrices weighted by number of streamlines and by mean streamline length between node pairs) for each pair of tractograms and node parcellation images. Subsequently, we used Brain Connectivity Toolbox (BCT)[3] to calculate modularity, average betweenness centrality, and global efficiency. We selected these graph theory metrics because they are statistically shown to be "streamline count invariant"[34], which guarantees their validity even though our harmonization methods don't preserve total streamline counts. Modularity is computed on connectivity matrices weighted by number of streamlines; global efficiency is derived from matrices comprising normalized streamline count; and average betweenness centrality is calculated on connectivity matrices weighted by mean streamline length. The evaluation of these graph measures is conducted prior to and following the three harmonization methods.

### 2.3. Harmonization methods

For calibrating connectivity matrices through mean shift, an average voxel-wise difference matrix from both datasets was calculated and added or subtracted from every connectivity matrix of a dataset. Negative values were converted to 0 in preparation for graph measure calculations. Specifically, suppose we're shifting site 1's connectivity matrices closer to those of site 2, the shifting transformation is:

$$S1[i,j]' = \max(0, (S1[i,j] - \overline{S1}[i,j] + \overline{S2}[i,j])), \quad (1)$$

where S1 denotes the connectivity matrix of a site 1 subject, and $\overline{S1}$ and $\overline{S2}$ are the element-wise means of matrices for all site 1 and site 2 subjects respectively.

We applied the Python (version 3.8) implementation of neuroCombat[21] to harmonize connectivity matrices. To reformat data for neuroCombat, we converted upper halves of connectivity matrices (including the main diagonal, i.e. entries (i, j) where i ≥ j) to columns in a 3570 x 168 dataframe in which each row represents a matrix entry and each column is a participant. We formed a batch id vector in which site 1 is encoded as 1 and site 2 is encoded 2. We further constructed a biological covariates matrix in which sex and age are supplied so that the variance attributable to those demographics can be retained in the data. After harmonization with neuroCombat, we reassembled the columns back to their symmetric matrix form, with negative values set to 0.

Lastly, we used the PyTorch (version 2.0.1) implementation of CycleGAN[23] released by the paper authors to style transfer the connectivity matrices from one dataset to another. Since the original model normalizes RGB values to [-1, 1] before training, we normalized connectivity matrices before inputting them to the model. We normalized matrices weighted by number of streamlines with this formula:

$$M[i,j]' = \frac{\ln(M[i,j])}{0.5 \times \ln(\max(M1, M2, \ldots))} - 1, \qquad (2)$$

where M represents a subject's connectivity matrices weighted by number of streamlines, and max(M1, M2,…) represents the largest element among all matrices of the two sites. We normalized matrices weighted by mean streamline length with this formula:

$$M[i,j]' = \frac{M[i,j]}{0.5 \times \max(M1, M2, \ldots)} - 1, \qquad (3)$$

where M represents a subject's connectivity matrices weighted by mean streamline length, and max(M1, M2,…) similarly represents the largest element among all matrices of the two sites. We modified the data loader and output format to take in and output one-channel NumPy array files. A 5-fold stratified cross-validation was implemented to train and test the model. We set load size as 100 and crop size as 84 to fit the shape of our arrays, set input channel and output channel as one, and disabled identity loss. Other parameters are set as default.

**2.4. Evaluation framework**

The effectiveness of the harmonization methods was evaluated based on two criteria: the removal of unwanted site effect and the preservation of between-subject biological variability. To assess the first criterion, we plotted the distributions of graph measures, computed combined datasets' CoV, and performed Mann-Whitney U test on datasets before and after applying different harmonization methods. A p-value greater than 0.05 in the Mann-Whitney U test indicated insignificant differences between distributions of the two sites' graph measures. This was regarded as successful removal of site effect. For the second criterion, the correlation between age and graph measures were analyzed. We considered a p-value less than 0.05 for the correlation coefficient to be a significant correlation between age and the graph measures.

## RESULTS

Before harmonization, the effect of site on graph measures is apparent (the leftmost upper and lower cells for each graph measure in Figure 3). Differences of global efficiency and modularity between the two datasets are especially substantial. To determine how effective mean shift and CycleGAN transfers the distributions of one dataset's graph measures to another (the six plots in the left of each cell) and how close distributions of the graph measures are after applying neuroCombat (the two plots in the right of each cell), we plotted values of the graph measures before and after the three harmonization methods. We observed that all harmonization methods worked to some degree for global efficiency and modularity measures: mean shift and CycleGAN approaches moved averages of one dataset's graph measures closer to another, with some successful reproductions of the overall distributions. For example, using CycleGAN to style transfer site 2's global efficiency yielded a difference of 0.0021 between the two sites' means. neuroCombat worked well, with differences difficult to distinguish between the processed datasets (global efficiency: 0.00019, modularity: 0.00015). However, all harmonization methods, except for mean shift in one direction, resulted in larger average disparities in betweenness centrality between the two sites. These results suggest that the harmonization methods work better in harmonizing connectivity matrices weighted by number of streamlines, from which we calculated global efficiency and modularity, compared with matrices weighted by mean streamline length, from which we derived average betweenness centrality.

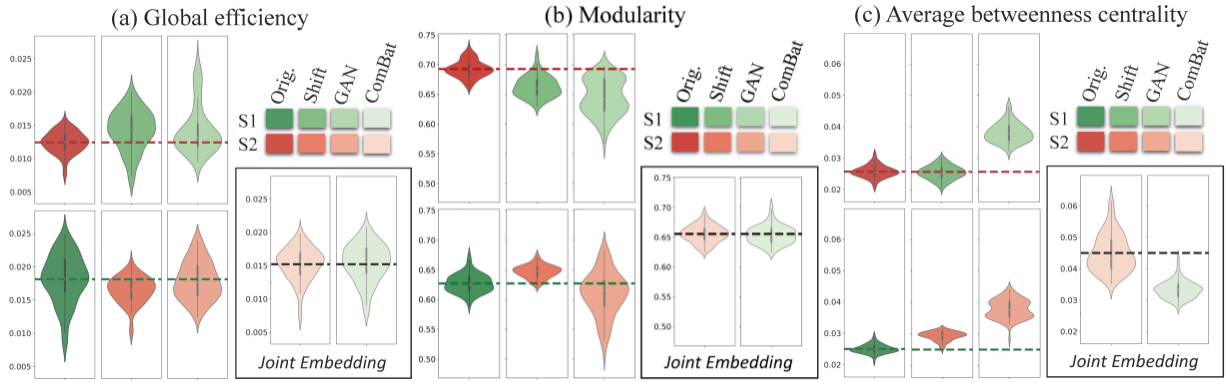

**Figure 3.** The three plots in each cell's top left show how well site 1 was normalized to site 2 using shift or GAN; the three plots in each cell's bottom left show how well site 2 was normalized to site 1 using shift or GAN; the tow plots in the bottom right (called "joint embedding") show how well neuroCombat harmonized the two sites. The dashed lines represent the baseline mean of the leftmost dataset in each group. The harmonization methods all worked to varying degrees on matrices weighted by number of streamlines degrees (neuroCombat better than CycleGAN and mean shift) but performed poorly on matrices weighted by mean streamline length.

We calculated the CoV of the joint dataset's graph measures (Figure 4) by dividing the standard deviations of the combined population by the corresponding mean. The basic trends supported our observations about Figure 3. CycleGAN's harmonization potential is questionable given its limited reduction of global efficiency's CoV. This is because the model generated results which fluctuated substantially, causing greater standard deviation.

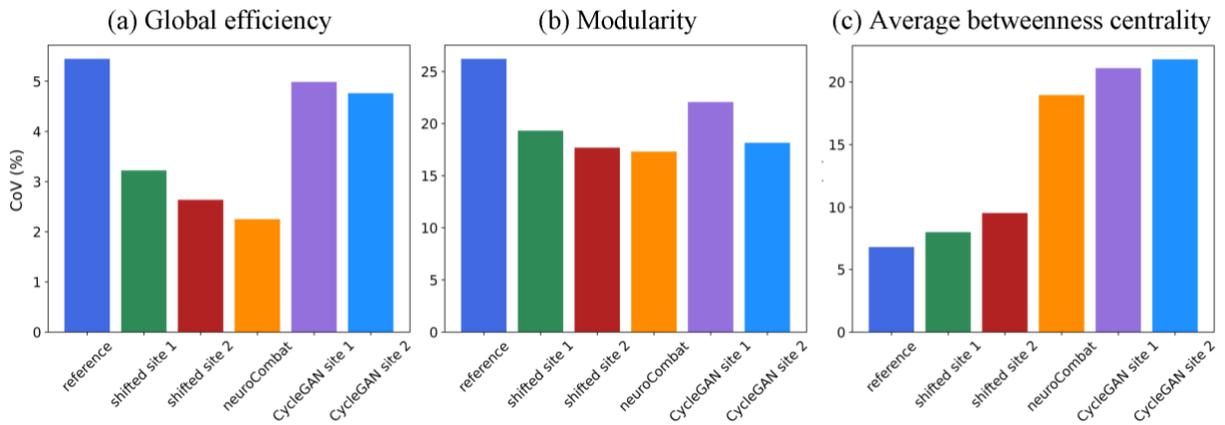

**Figure 4.** A reduction in CoV indicates effective harmonization. From the plots of global efficiency and modularity, both mean shift and neuroCombat effectively reduced CoV. Meanwhile, CycleGAN was less effective – note the purple and the light blue. For average betweenness centrality, no method was particularly effective. The least harmful method was mean shift. neuroCombat and CycleGAN nearly tripled the CoV.

Prior research has demonstrated significant negative association between global efficiency and age as well as significant positive correlations between modularity and age.[35][7] No work demonstrated a statistically significant correlation between age and average betweenness centrality of dMRI connectivity matrices. We performed a linear regression analysis to evaluate whether our harmonization methods preserve or strengthen these trends.

Before harmonization, neither modularity nor global efficiency indicates a significant correlation with age (Table 4). All harmonization methods except for style transferring site 2 with CycleGAN make the correlations between age and the two metrics more significant. Mean shifting site 1 attains the most significant correlations among all methods with p-values less than 0.001 for both metrics. CycleGAN in different style transferring directions generated regression lines with the steepest and flattest slope (Figure 5) for global efficiency and modularity. No significant correlation between age and average betweenness centrality was found before or after harmonization.

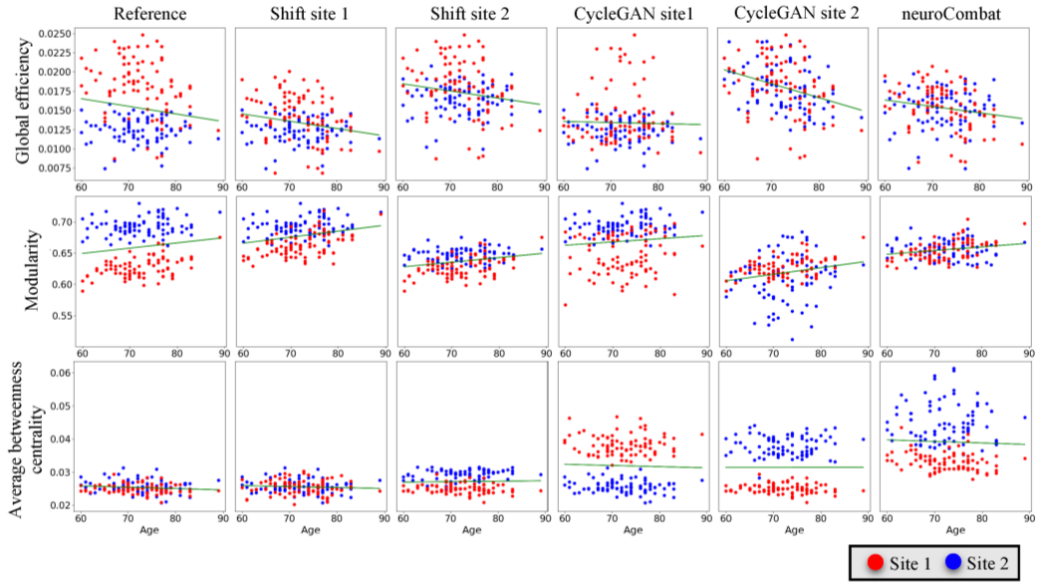

**Figure 5.** All harmonization methods preserve the correlations between age and global efficiency and between age and modularity to varying degrees. CycleGAN generates the steepest correlation for global efficiency and modularity when style transferring site 2 and the flattest correlation when style transferring site 1.

**Table 4.** Mean shifting site 1 leads to the most significant correlation between age and modularity with p-value smaller than 0.001 for either direction. Style transferring site 1 with CycleGAN keeps the correlations insignificant. Other methods generally make the correlations between age and global efficiency and between age and modularity more significant. No significant correlation was found between age and average betweenness centrality before or after harmonization.

|  | Reference | Mean shift site 1 | Mean shift site 2 | CycleGAN on site 1 | CycleGAN on site 2 | neuro-Combat |
|---|---|---|---|---|---|---|
| Pearson's r between age and global efficiency | -0.148 | -0.221*** | -0.181* | -0.028 | -0.331*** | -0.188* |
| Pearson's r between age and modularity | 0.140 | 0.264*** | 0.261*** | 0.094 | 0.213* | 0.239** |
| Pearson's r between age and average betweenness centrality | -0.1443 | -0.0848 | 0.0315 | -0.0314 | 0.0001 | -0.0377 |

"*" indicates p<0.05, "**" indicates p<0.005, and "***" indicates p<0.001.

We applied the Mann-Whitney U test to quantify the differences between the derived graph measures' distributions of two sites (Figure 5). Specifically, each cell in Table 5 measures the differences between the blue and the red dots in the corresponding scatter plot in Figure 5 by ranking the dots of the combined dataset. A significant p-value resulting from the test implies that site plays a substantial role in influencing the distribution of the respective graph measure.

Differences between the two sites are statistically significant for all graph measures with p-value less than 0.005 before harmonization (Table 4). No harmonization method successfully removed the significant site differences for all graph measures. As we observed previously, neuroCombat performed well for harmonizing connectivity matrices weighted by number of streamlines. No significant site effect on global efficiency and modularity was observed after applying neuroCombat. Mean shifting site 1 achieved the best result for harmonizing connectivity matrices weighted by mean streamline length, with no significant effect of site on average betweenness centrality after harmonizing with this method. The statistical tests suggest that though most methods harmonized the connectivity matrices weighted by number of streamlines in the right directions (Figure 4), many failed to completely remove site effects (Table 5).

**Table 5.** Results of the Mann-Whitney U test investigating the effect of site on graph measures. The null hypothesis is that there is no difference between the distributions of the two independent groups being tested. No asterisk indicates that the differences between the two groups are statistically insignificant. The smaller the p-value, the more likely that there are observed differences between datasets to reject the null hypothesis. No harmonization method performs consistently well across all metrics.

|  | Reference | Shift site 1 | Shift site 2 | CycleGAN site 1 | CycleGAN site 2 | neuro-Combat |
|---|---|---|---|---|---|---|
| Global efficiency | 611.0*** | 1976.0*** | 2306.5*** | 2585.0 | 3234.5** | 3338.5 |
| Modularity | 7043.0*** | 6388.0*** | 6055.0*** | 6552.0*** | 2424.0*** | 3815.5 |
| Average Betweenness Centrality | 4419.5** | 3726.5 | 6869.0*** | 7054.0*** | 6941.0*** | 0.0*** |

"*" indicates p<0.05, "**" indicates p<0.005, and "***" indicates p<0.001.

## DISCUSSION

This study shows the difficulties of harmonizing at the connectivity matrix level, though certain methods perform well for certain metrics. This is in line with previous finding which indicates that harmonization of parameters yields superior results compared with harmonization of matrices.[19]

Mean shift strengthens within-site biological variability but reduces the least site effects on graph measures. Transforming negative values to zeros after shifting means and not considering standard deviations are the method's limitations.

neuroCombat performs relatively well for global efficiency and modularity. The method removes significant site effect on graph measures and strengthens the correlations between age and these two metrics to significant. However, as other methods, based on the nearly tripled CoV, ComBat is ineffective in harmonizing average betweenness centrality.

CycleGAN generates visually harmonized results, and the model performs the best for certain direction and metric (e.g. style transfer site 1 to site 2, global efficiency). However, the model is not stable, generating results which neither preserve biological variability nor remove much site effect under other conditions. This could happen due to the relatively small datasets in this study (84 images in each dataset), natural log normalization not preserving precision of values, and inputs of low-resolution array (84 by 84 matrix) but default number of layers in CycleGAN as a smaller generator and discriminator might work better for low-resolution image/array. Since CycleGAN is typically applied on images, the different spatial structure of connectivity matrices could also contribute to the poor performance. Future studies could try to address these limitations and examine the performance of fine-tuned CycleGAN on harmonizing connectivity matrices.

A shared problem of all these harmonization methods is the failure to harmonize matrices weighted by mean streamline length. Before harmonization, the two datasets already have close distributions of average betweenness centrality, graph measure calculated from connectivity matrices weighted by mean streamline length. However, all harmonization methods increased the differences in average betweenness centrality between the two sites. Study indicates the irreproducibility of betweenness centrality.[36] However, an examination of characteristic path length, another graph metric derived from connectivity matrices weighted by mean streamline length, before and after the harmonization methods shows similar ineffectiveness. We chose not to include this in our study due to the metric's strong dependence on streamline count.[34] Direct comparison of its value before and after harmonization could be problematic, as our methods don't preserve total streamline count. Further experiments could be done on other metrics such as small-worldness and randomness to verify the harmonization methods' limited effectiveness when applied on matrices weighted by mean streamline length.

This work enhances our understanding of connectivity matrix space and assesses the performance of mean shift, neuroCombat, and CycleGAN on harmonizing multi-site connectivity matrices. Future studies could consider reproducing the results of this research on different datasets and acquisition methods, brain parcellation methods, tractography methods, types of values connectivity matrices weighted by, and graph measures. Experiments of these methods on data from more than two sites are valuable. Testing if the harmonization method prevents a classifier from identifying the site of origin of a given sample would provide a further evaluation of the method's effectiveness.

## ACKOWDGEMENTS

This work was supported by the National Institutes of Health (NIH) under award numbers K01EB032989, K01EB032898, K24-AG046373, K01-AG073584, and R01-AG034962, the National Science Foundation (NSF) under award number 2040462, the Alzheimer's Association under award IIRG-08-88733, the Vanderbilt Clinical Translational Science Awards UL1-TR000445 and UL1-TR002243, Vanderbilt's High-Performance Computer Cluster for Biomedical Research under award S10-OD023680, and the Integrated Training in Engineering and Diabetes grant number T32 DK101003. This work was conducted in part using the resources of the Advanced Computing Center for Research and Education at Vanderbilt University, Nashville, TN. This research was conducted with the support from the Intramural Research Program of the National Institute on Aging of the NIH. The Vanderbilt Institute for Clinical and Translational Research (VICTR) is funded by the National Center for Advancing Translational Sciences (NCATS) Clinical Translational Science Award (CTSA) Program, Award Number 5UL1TR002243-03. The content is solely the responsibility of the authors and does not necessarily represent the official views of the NIH or NSF.

Study data were obtained from the Vanderbilt Memory and Aging Project (VMAP). VMAP data were collected by Vanderbilt Memory and Alzheimer's Center Investigators at Vanderbilt University Medical Center. This work was supported by NIA grants R01-AG034962 (PI: Jefferson), R01-AG056534 (PI: Jefferson), R01-AG062826 (PI: Gifford), Alzheimer's Association IIRG-08-88733 (PI: Jefferson), and R01-EB017230 Controlling Quality and Capturing Uncertainty in Advanced Diffusion Weighted MRI. The BIOCARD study is supported by a grant from the National Institute on Aging (NIA): U19-AG03365. The BIOCARD Study consists of 7 Cores and 2 projects with the following members: (1) The Administrative Core (Marilyn Albert, Corinne Pettigrew, Barbara Rodzon); (2) the Clinical Core (Marilyn Albert, Anja Soldan, Rebecca Gottesman, Corinne Pettigrew, Leonie Farrington, Maura Grega, Gay Rudow, Rostislav Brichko, Scott Rudow, Jules Giles, Ned Sacktor); (3) the Imaging Core (Michael Miller, Susumu Mori, Anthony Kolasny, Hanzhang Lu, Kenichi Oishi, Tilak Ratnanather, Peter vanZijl, Laurent Younes); (4) the Biospecimen Core (Abhay Moghekar, Jacqueline Darrow, Alexandria Lewis, Richard O'Brien); (5) the Informatics Core (Roberta Scherer, Ann Ervin, David Shade, Jennifer Jones, Hamadou Coulibaly, Kathy Moser, Courtney Potter); the (6) Biostatistics Core (Mei-Cheng Wang, Yuxin Zhu, Jiangxia Wang); (7) the Neuropathology Core (Juan Troncoso, David Nauen, Olga Pletnikova, Karen Fisher); (8) Project 1 (Paul Worley, Jeremy Walston, Mei-Fang Xiao), and (9) Project 2 (Mei-Cheng Wang, Yifei Sun, Yanxun Xu).